\documentclass[a4paper,11pt]{article}
\usepackage{pos}
\pdfoutput=1 
\usepackage{lineno}
\usepackage{ptdr-definitions}
\usepackage{hepunits}
\usepackage{heppennames2}
\usepackage{xpatch}
\makeatletter
\xpatchcmd\@HepConStyle
 {\edef\@upcode{\updefault}}
 {\ifdefined\shapedefault\edef\@upcode{\shapedefault}\else\edef\@upcode{\updefault}\fi}
 {}{}
\makeatother
\title{Heavy flavor collectivity in small systems}
\author*[a]{Georgios Konstantinos Krintiras}
\affiliation[a]{The University of Kansas\\ \href{https://gkrintir.web.cern.ch/}{cern.ch/gkrintir}}

\emailAdd{gkrintir@cern.ch}

\abstract{
The presence of correlations between particles significantly separated in pseudorapidity in proton-proton and proton-nucleus collisions revealed surprises in the early LHC data. Are the physical processes responsible for the observed long-range pseudorapidity correlations and their azimuthal structure the same in small collision systems as in heavy ion collisions? Whereas in the case of heavy ion collisions ``flow'' is interpreted as generated by initial geometric inhomogeneities, calculations indicate that initial-state momentum correlations are present and could contribute to the observed azimuthal anisotropy in small systems. Probes involving heavy quarks provide us with a unique opportunity to disentangle different quantum chromodynamics effects at the boundary between low- and high-\pt interactions, and hence shed light on the origin of flow in small collision systems. A selection of the latest measurements is presented for the flow and production of heavy flavor hadrons and their decay products.
}

\FullConference{%
 
 The Ninth Annual Conference on Large Hadron Collider Physics - LHCP2021\\
 7--12 June 2021\\
 Online
}

\begin{document}
\maketitle


\section{Introduction}

Extensive measurements of light hadron azimuthal anisotropies (characterized by
the Fourier coefficients $v_n$) have been performed in all collision systems at LHC~\cite{Krintiras:2021koj}, indicating the formation and hydrodynamic expansion (``flow'') of a region of hot and dense quark–gluon plasma (QGP) with a small ratio of the shear viscosity to entropy density. Heavy flavor (charm and bottom) quarks have masses much larger than the typical range of temperatures in the QGP, meaning thermal production of heavy quarks during the QGP phase is suppressed relative to that of light quarks. 

However, it is still postulated that charm quarks interact strongly enough to flow with the QGP. Experimental data at LHC~\cite{ALICE:2008ngc,Aad:2008zzm,Chatrchyan:2008zzk} indeed reveal that charm quark hadrons, as well as their decay leptons, have significant azimuthal anisotropies, suggesting that they participate in the overall collective flow of the medium (see Section~\ref{sec:HF_AA}). A nonthermalized probe is required to assess the interaction with the medium more thoroughly, with the heavier bottom quark being a natural candidate. Although a series of theoretical predictions for the azimuthal anisotropies of bottom quarks exist, only limited experimental data are currently available at LHC (see Section~\ref{sec:HF_AA}). 

Smaller collision systems, including proton-nucleus ($\Pp{}\mathrm{Pb}$) and even proton-proton ($\Pp{}\Pp$) collisions, have particle emission patterns with large azimuthal anisotropies that can be also described by nearly inviscid hydrodynamics. It is of interest to measure heavy flavor anisotropies in such collision systems so that we obtained information about the interaction of heavy quarks with the medium in the smallest hadronic collision systems at LHC (see Section~\ref{sec:HF_pA_pp}). The measurement of the relative production of different heavy flavor hadron species is also sensitive to the charm- and beauty-quark fragmentation and hadron formation processes (see Section~\ref{sec:HF_other}).

\section{Heavy flavor flow measurements in heavy ion collisions}
\label{sec:HF_AA}

\begin{figure}[!htb]
\centering
\begin{minipage}{0.5\textwidth}
\centering
\includegraphics[width=0.85\textwidth]{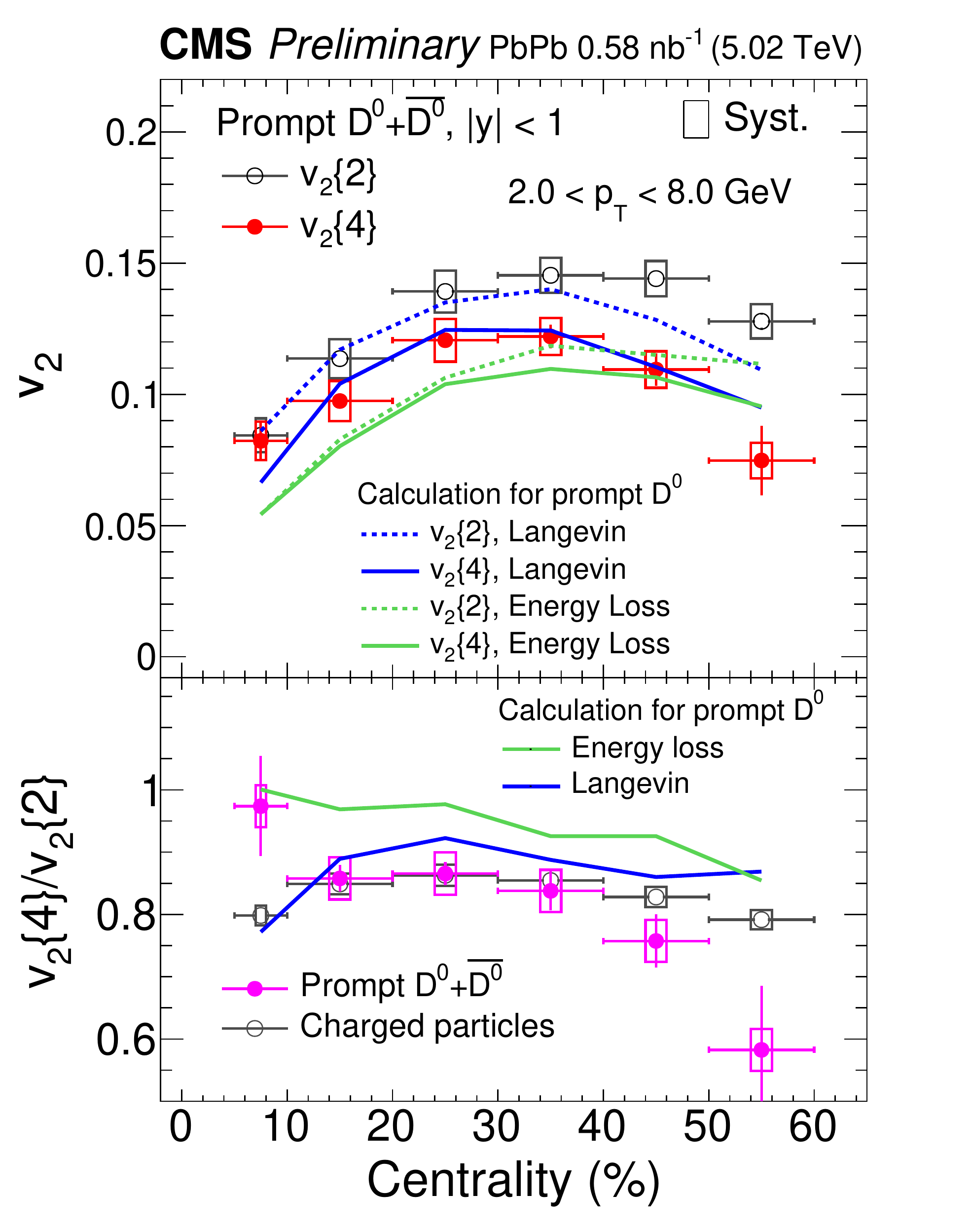}
\end{minipage}\hfill
\begin{minipage}{0.5\textwidth}
\centering
\includegraphics[width=0.99\textwidth]{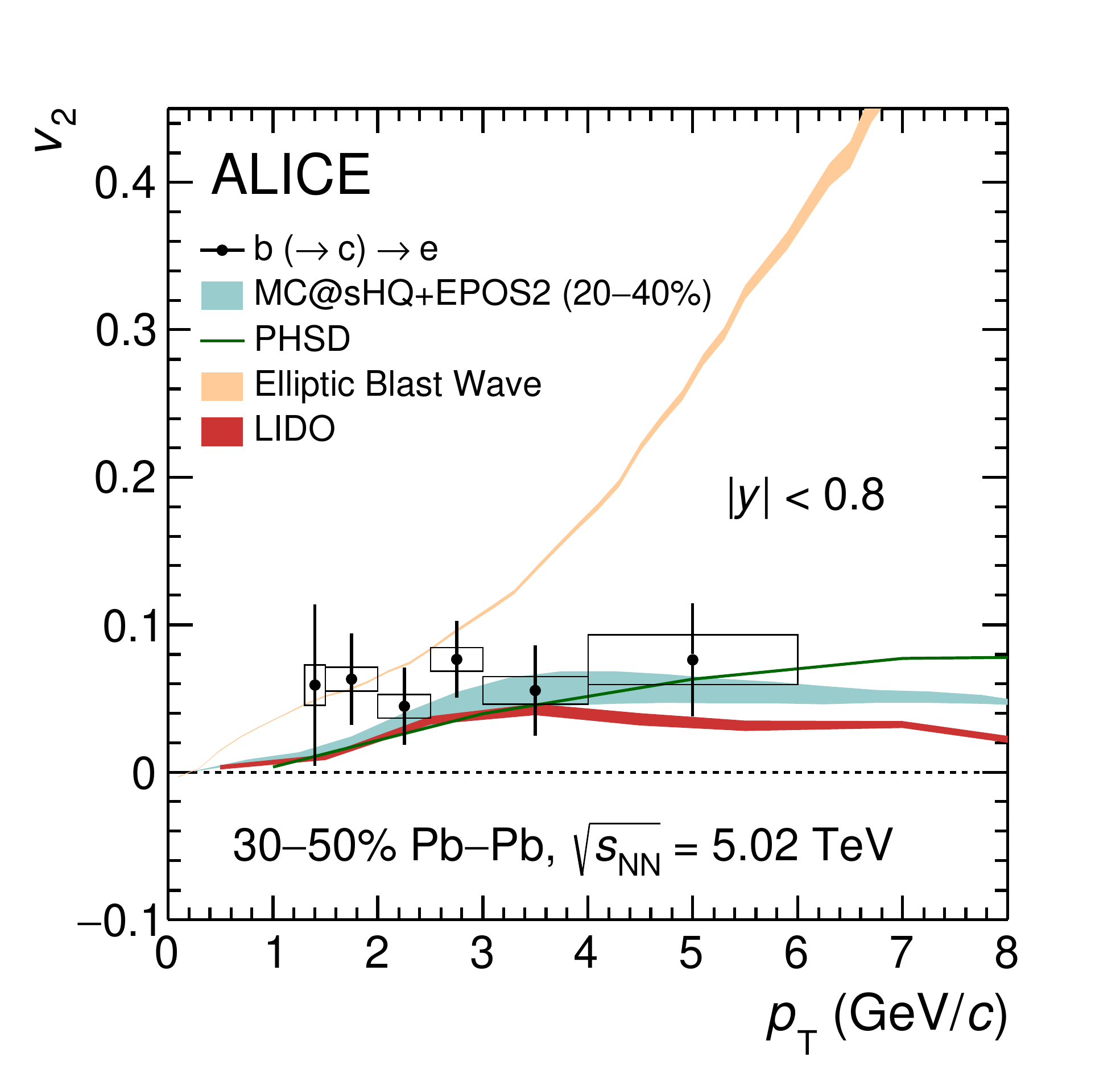}
\end{minipage}
\caption{Left: Prompt {\PDz} meson $v_2$ (upper panel) and $v_2$ ratios compared to those for charged particles (lower panel) as a function of the centrality class from $\mathrm{Pb}{}\mathrm{Pb}$ collisions at $\sqrt{\smash[b]{s_{_{\mathrm{NN}}}}}$ = 5.02\,TeV. The lines denote model calculations~\protect\cite{CMS-PAS-HIN-20-001}. Right: Elliptic flow of electrons from bottom hadron decays in the 30--50\% centrality class in $\mathrm{Pb}{}\mathrm{Pb}$ collisions at $\sqrt{\smash[b]{s_{_{\mathrm{NN}}}}}$ = 5.02\,TeV as function of \pt compared with model calculations~\protect\cite{ALICE:2020hdw}.}
\label{fig:HF_AA}
\end{figure}

At LHC, a significant $v_2$ (``elliptic'') flow signal is observed for mesons containing a charm quark, \eg, prompt $\PJGy$~\cite{Khachatryan:2016ypw,ATLAS:2018xms,ALICE:2020pvw} and prompt $\PDz$~\cite{ALICE:2017pbx,Sirunyan:2017plt}, while the first measurements with bottom quarks, \eg, nonprompt $\PJGy$~\cite{CMS:2016mah,ATLAS:2018xms}, and $\PGU$(1S)~\cite{ALICE:2019pox,CMS:2020efs} and $\PGU$(2S)~\cite{CMS:2020efs}, are compatible with zero in the kinematic region studied so far. For the charm quark case, the prompt $\PDz$ meson $v_2$ has been so far measured
using two-particle correlation methods, while the first $v_2\{4\}$ measurement, \ie, using
multiparticle correlations, is presented most recently~\cite{CMS-PAS-HIN-20-001} (Fig.~\ref{fig:HF_AA}, left). For the bottom quark case, measurements of the elliptic and ``triangular'' ($v_3$) flow of electrons~\cite{ALICE:2020hdw} (Fig.~\ref{fig:HF_AA}, right) and muons~\cite{ATLAS:2020yxw} originating from beauty hadron decays are also important for the understanding of the degree of thermalization of beauty quarks in the QGP. At high \pt, the results can significantly discriminate between models of heavy quark energy loss and constrain heavy quark transport coefficients in the QGP, complementing measurements of the nuclear modification factor, \eg, Ref.~\cite{CMS:2018bwt}.

\section{Heavy flavor flow measurements in small systems}
\label{sec:HF_pA_pp}

\begin{figure}[!htb]
\centering
\begin{minipage}{0.5\textwidth}
\centering
\includegraphics[width=0.85\textwidth]{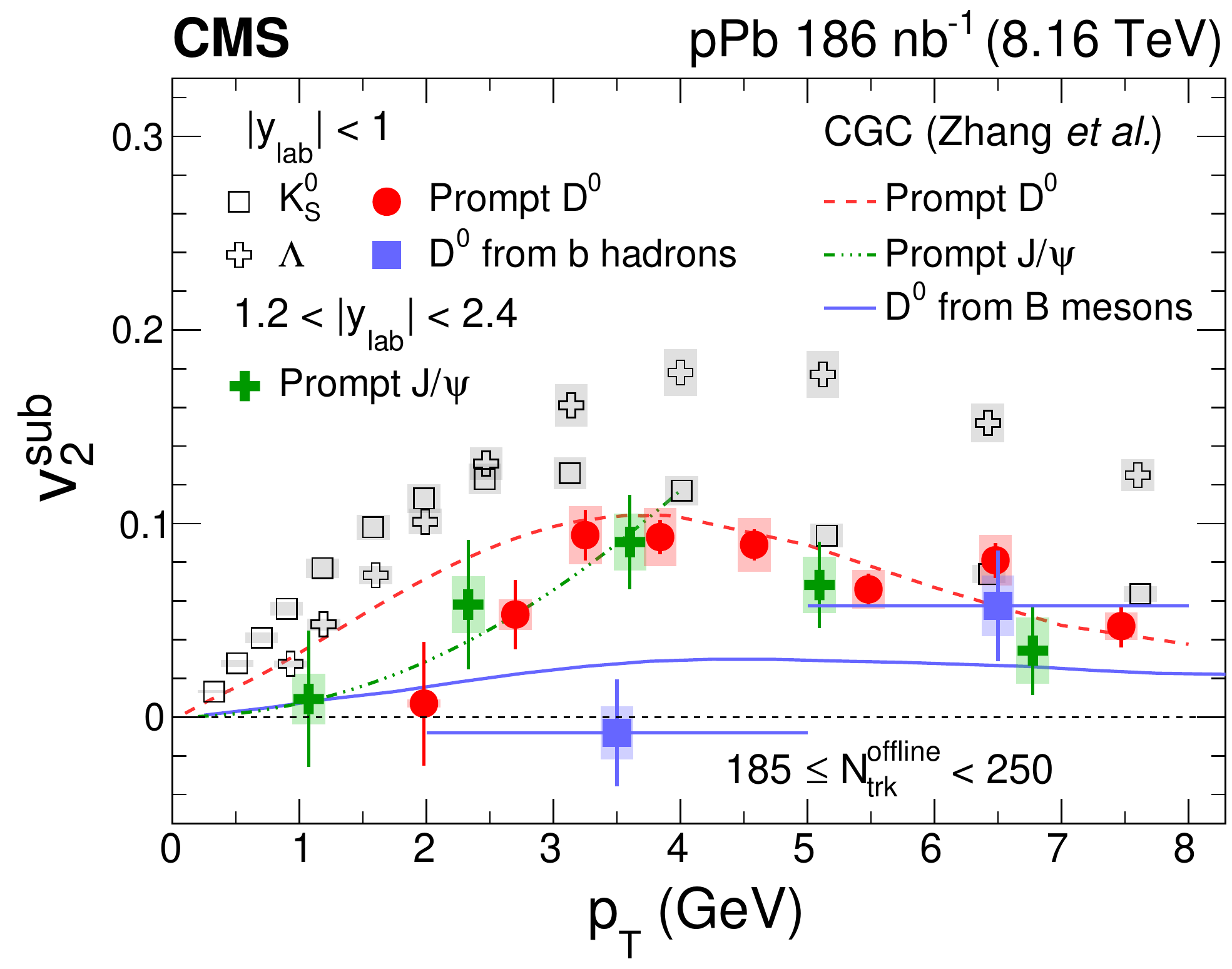}
\end{minipage}\hfill
\begin{minipage}{0.5\textwidth}
\centering
\includegraphics[width=0.99\textwidth]{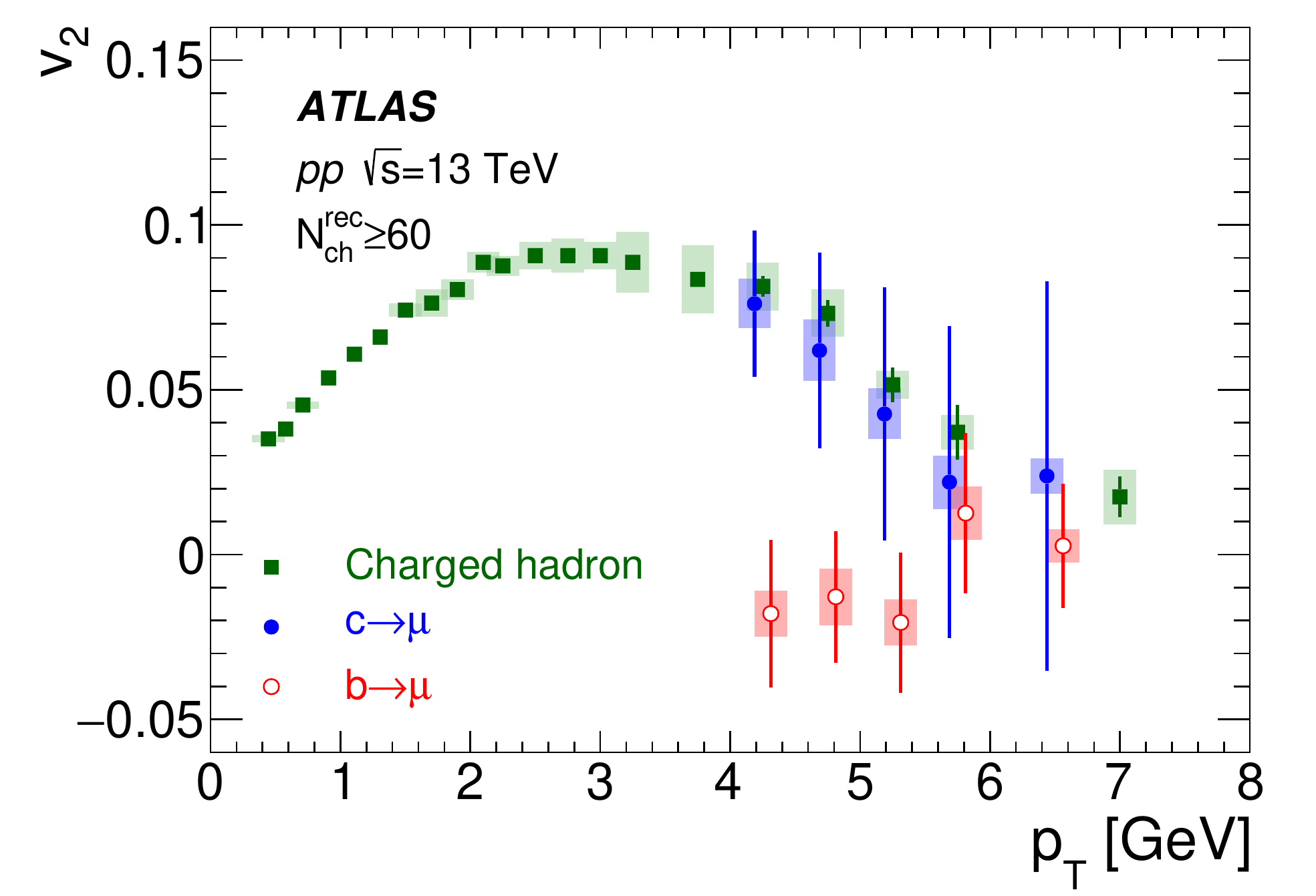}
\end{minipage}
\caption{Left: Results of $v_2$ for prompt and nonprompt \PDz mesons, as well as prompt \PJGy\ mesons and light flavor hadrons, as a function of \pt in $\Pp{}\mathrm{Pb}$ collisions at $\sqrt{\smash[b]{s_{_{\mathrm{NN}}}}}= 8.16$\,TeV. Lines show the theoretical calculations of prompt and nonprompt \PDz, and prompt \PJGy mesons, respectively~\protect\cite{CMS:2020qul}. Right: Comparison of $v_2$ between charged hadrons and muons originating from charm or bottom hadrons as a function of \pt in $\Pp{}\Pp$ collisions at 13\,TeV~\protect\cite{ATLAS:2019xqc}. For the case of heavy flavor muons, there are smearing effects due to decay kinematics relative to the parent hadron, so caution is warranted when comparing hadron and heavy flavor muon flow.}
\label{fig:HF_pA_pp}
\end{figure}

In small colliding systems, the study of heavy flavor hadron collectivity has the potential to disentangle possible contributions from both initial- and final-state effects.  Recent observations of significant $v_2$ signal for prompt \PDz~\cite{ALICE:2017pbx,CMS:2018loe} and prompt \PJGy~\cite{ALICE:2017smo,CMS:2018duw} in $\Pp{}\mathrm{Pb}$ collisions (Fig.~\ref{fig:HF_pA_pp}, left) provided the first evidence for charm quark collectivity in small systems. In spite of the mass differences, the observed $v_2$ signal for prompt \PJGy mesons is found to be comparable to that of prompt \PDz mesons and light-flavor hadrons at a given \pt range, possibly implying the existence of
initial-state correlation effects. 

Further detailed investigations start addressing open questions for understanding the origin of heavy flavor quark collectivity in small
systems. These include the \pt and multiplicity dependence of charm quark collectivity in both $\Pp{}\mathrm{Pb}$ and $\Pp{}\Pp$ systems (Fig.~\ref{fig:HF_pA_pp}, right), and the details of collective behavior of beauty quarks in the $\Pp{}\mathrm{Pb}$ system~\cite{CMS:2020qul}. Additionally, heavy flavor decay electrons~\cite{ALICE:2018gyx} or muons~\cite{ATLAS:2019xqc} provide information about the anisotropies of leptons from charm and bottom hadron decays in a combined or separate way.

\section{Other measurements}
\label{sec:HF_other}

\begin{figure}[!htb]
\centering
\begin{minipage}{0.5\textwidth}
\centering
\includegraphics[width=0.85\textwidth]{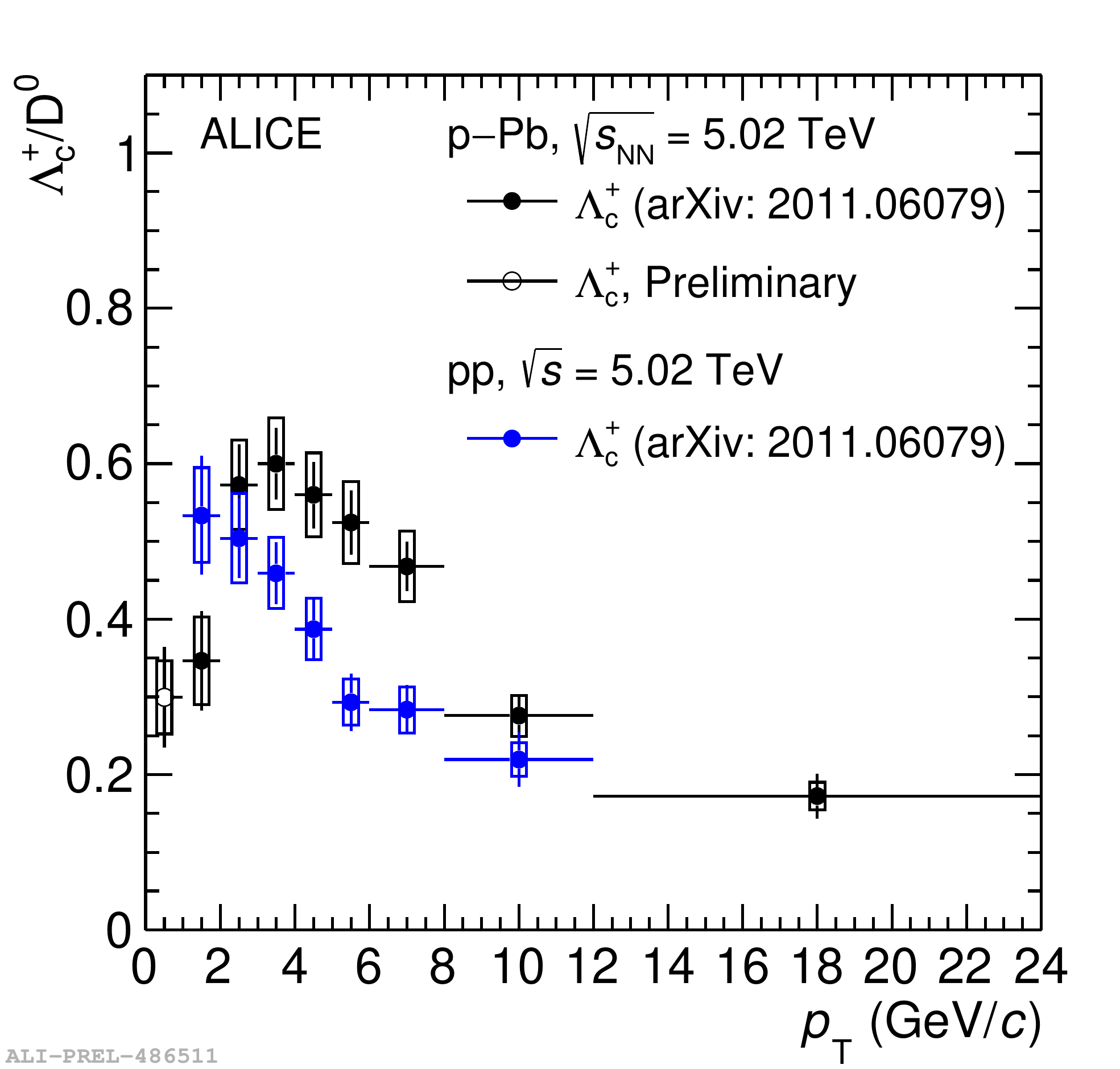}
\end{minipage}\hfill
\begin{minipage}{0.5\textwidth}
\centering
\includegraphics[width=0.99\textwidth]{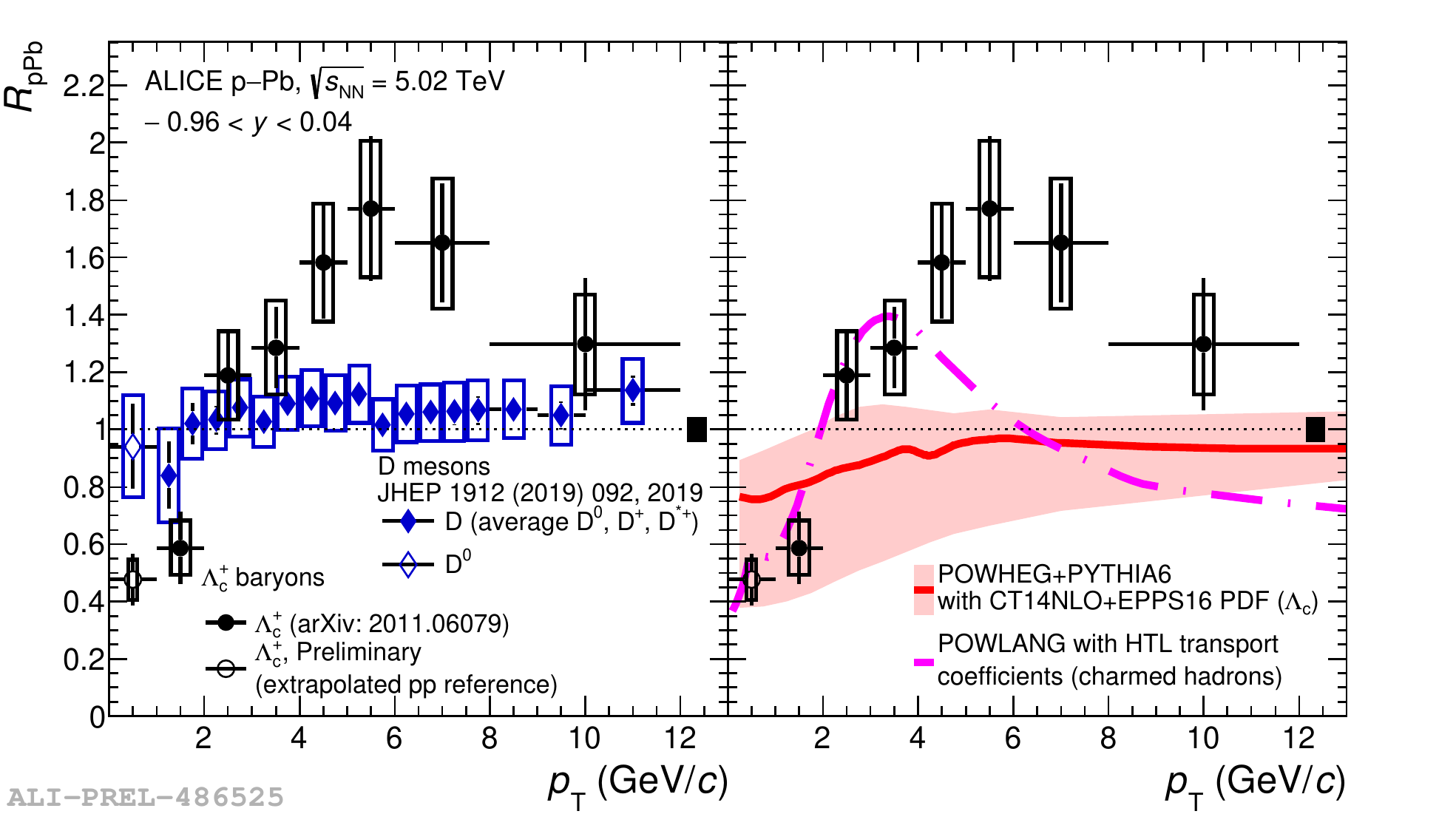}
\end{minipage}
\caption{Left: The $\PGLpc/\PDz$ ratio as a function of \pt measured in $\Pp{}\mathrm{Pb}$ and $\Pp{}\Pp$ collisions~\protect\cite{ALICE:2020wla}. Right: The nuclear modification factor of prompt \PGLpc baryons in $\Pp{}\mathrm{Pb}$ collisions as a function of \pt, compared to \PDz mesons, as well as model calculations assuming that a hot deconfined medium is formed~\protect\cite{ALICE:2020wla}. }
\label{fig:HF_other}
\end{figure}

The measurement of charm and beauty hadron production is a powerful test of perturbative quantum chromodynamics calculations. The $\PGLpc/\PDz$ baryon-to-meson ratio in $\Pp{}\Pp$ and $\Pp{}\mathrm{Pb}$ collisions~\cite{CMS:2019uws,ALICE:2020wla}, similar to ratios measured in the light flavor sector, is larger than measurements obtained in elementary collision systems at lower center-of-mass energies, challenging the assumption that fragmentation fractions for quarks are universal. The increase of LHC Run 2 (2015--2018) data precision even allowed for the observation of a clear decreasing trend down to $\pt=0$\,GeV in the $\PGLpc/\PDz$ ratio for the $\Pp{}\mathrm{Pb}$ case~\cite{ALICE:2020wla} (Fig.~\ref{fig:HF_other}, left). Additionally, the nuclear modification factor of \PGLpc~\cite{ALICE:2020wla}
 baryons is found to be higher than \PD mesons in the intermediate \pt region, although the current precision is not enough to draw conclusions on the role of different effects, \eg, the possible presence of hot-medium effects (Fig.~\ref{fig:HF_other}, right).

\section{Summary}

Heavy quarks are typically produced early in the collisions via high-momentum transfers between incoming partons. Once created, heavy quarks persist throughout the dynamical time evolution of the quark-gluon plasma, and hence serving
as sensitive probes of the hot and dense medium. Although a common hydrodynamic
description of azimuthal anisotropies in all colliding systems as resulting from initial geometry anisotropies is compelling, heavy flavor hadrons could discriminate against possible initial-state effects, \eg, gluon saturation. The increased precision of measurements with respect to those performed with the LHC Run 2 (2015--2018) data is crucial for providing further insight into heavy flavor hadron production in proton-nucleus and proton-proton collisions~\cite{Citron:2018lsq,ALICE:2020fuk}.

\bibliographystyle{auto_generated}
\bibliography{skeleton}

\end{document}